\documentclass[preprint,3p,12pt]{elsarticle}

\usepackage{empheq}
\usepackage{widetext}

\usepackage{amssymb}
\usepackage{amsmath}
\usepackage[tight-spacing=true]{siunitx}

\usepackage{color}
\usepackage{url}
\usepackage{lscape}
\usepackage{multirow}
\usepackage{pdflscape}
\usepackage{dcolumn}  
\usepackage[colorlinks = true,
            linkcolor = blue,
            urlcolor  = blue,
            citecolor = blue,
            anchorcolor = blue]{hyperref}

\DeclareSIUnit\wn{\cm\tothe{-1}}
\DeclareSIUnit\bar{bar}

\journal{J. Mol. Spectrosc.}

\begin{document}

\begin{frontmatter}



\title{\texorpdfstring{High-resolution far-infrared synchrotron FTIR spectroscopy and analysis of the $\nu_7$, $\nu_{19}$ and $\nu_{20}$ bands of trioxane}{High-resolution far-infrared synchrotron FTIR spectroscopy and analysis of the v7, v19 and v20 bands of trioxane}}


\author[icb]{C.~Richard\corref{cor1}}\ead{Cyril.Richard@u-bourgogne.fr}
\author[mon]{P.~Asselin}
\author[icb]{V.~Boudon}

\address[icb]{Laboratoire Interdisciplinaire Carnot de Bourgogne, UMR 6303 CNRS - Universit\'e Bourgogne Franche-Comt\'e, 9 Av. A. Savary, BP 47870, F-21078 Dijon Cedex, France}
\address[mon]{CNRS, De la Mol\'ecule aux Nano-Objets: R\'eactivit\'e, Interactions, Spectroscopies, MONARIS, Sorbonne Universit\'e, Paris, France.}

\begin{abstract}
Rovibrational band spectra of the three $\nu_{20}$, $\nu_7$ and $\nu_{19}$ bands of 1, 3, 5 -- trioxane (H$_2$CO)$_3$ were recorded in the 50--\SI{650}{\wn} range using a long path absorption cell coupled to a high resolution Fourier transform spectrometer and synchrotron radiation at the AILES beamline of the SOLEIL synchrotron. More than 16\,000 lines were assigned with a dRMS better than $0.17\times10^{-3}$\si{\wn}. Two different formalisms (tensorial and Watson) were used to derive accurate rotational and quartic parameters for the three bands and for the first time. A precise determination of Coriolis parameter and $q_+$ $l-$doubling constant for both $\nu_{20}$ and $\nu_{19}$ perpendicular bands was also obtained. Lastly, each set of spectroscopic parameters is compared and discussed between both formalisms.
\end{abstract}

\begin{graphicalabstract}
\includegraphics[width=0.9\textwidth]{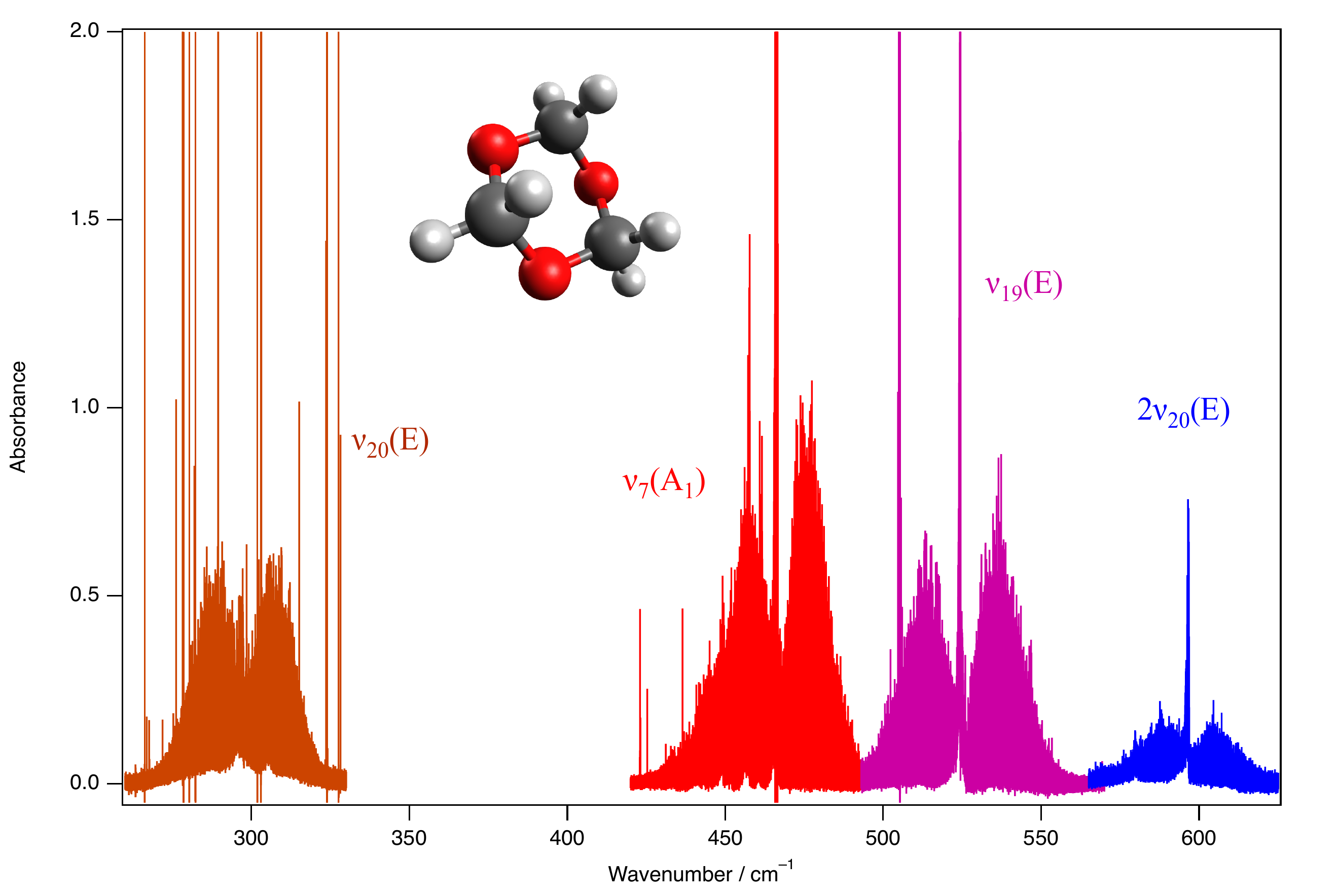}
\end{graphicalabstract}

\begin{highlights}
\item Far infrared high resolution spectroscopy of four bands of the trioxane
\item Complete line position analysis of $\nu_7$, $\nu_{19}$ and $\nu_{20}$ bands of trioxane
\item Comparison of tensorial and Watson's formalism results
\end{highlights}


\end{frontmatter}




\section{Introduction}
\begin{figure}[ht]
\centerline{\includegraphics[width=0.5\textwidth]{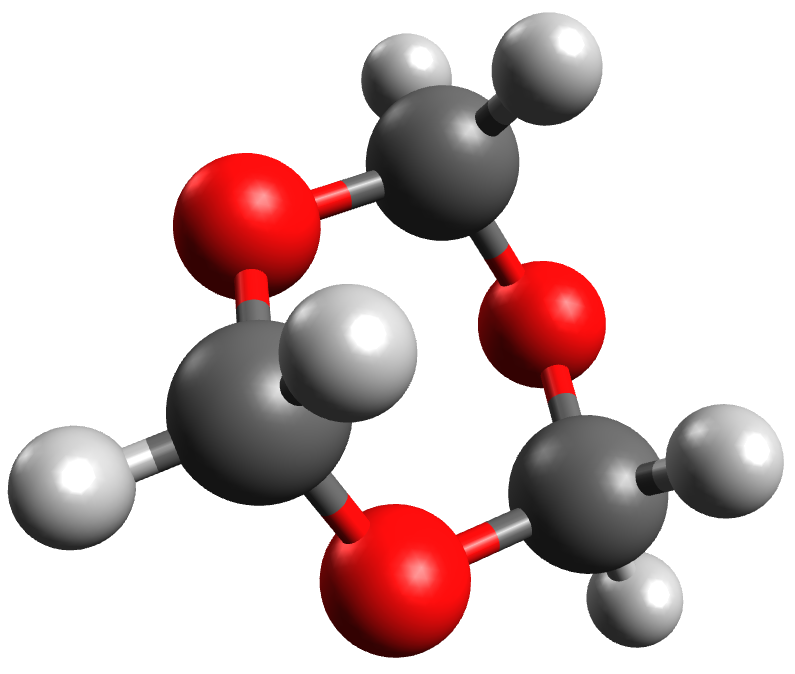}}
\caption{Three-dimensional representation of the trioxane molecule (H$_2$CO)$_3$.}
\label{fig:trioxane}
\end{figure}

The molecule of 1, 3, 5 -- trioxane (H$_2$CO)$_3$, a cyclic trimer of formaldehyde, is an oblate symmetric-top that belongs to the $C_{3v}$ symmetry group as illustrated in the Fig.~\ref{fig:trioxane}. It is a good example of a reasonably rigid molecule with 20 fundamental modes: 7 symmetric vibrations of type $A_1$ (parallel bands), 3 vibrations of type $A_2$ (for which absorption from the ground state
is symmetry-forbidden)) and 10 doubly degenerate vibrations of type $E$ (perpendicular bands).

Because of its relatively high number of atoms (12) and its fairly high mass, the rotational spectrum of trioxane is dense, leading to a possible radioastronomy detection. For many years, trioxane is also known as a molecule that could be detected in comet comae\cite{cottin2000experimental,cottin2004origin}, making it as highly relevant to studies of prebiotic chemistry. Its first microwave spectrum was obtained by Oka et \emph{al}.\cite{oka1964microwave} in 1963, and the analysis was extended by Colmont and co-workers\cite{bellet1970millimeter,colmont1975microwave,henninot1992infrared,colmont1980assignment} for the excited states below \SI{850}{\wn}. Submillimetric spectra were then measured in its ground-state and its two lowest lying excited states $\nu_7 = 1$ at \SI{467}{\wn} and $\nu_{20} = 1$ at \SI{307}{\wn}\cite{gadhi1989submillimeter}. Two jet-cooled mid-infrared spectroscopic studies reported rovibrational spectra of the $\nu_{17}$ and $\nu_{16}$ bands, respectively centered at 1071 and \SI{1177}{\wn}\cite{gibson2015rotationally}.

This paper presents a complete analysis of three bands, observed at high resolution in the far infrared: 
\begin{itemize}
 \item $\nu_{20}$ mode, CH$_2$ torsion, $E$, \SI{297}{\wn},
 \item $\nu_{7}$ mode, OCO bending, $A_1$, \SI{466}{\wn},
 \item $\nu_{19}$ mode, OCO bending, $E$, \SI{525}{\wn}.
\end{itemize}

Infrared spectra of trioxane have been recorded in the 50--\SI{650}{\wn} range using a high resolution Bruker IFS 125 interferometer located at the AILES beamline of the SOLEIL synchrotron facility. Owing to its higher brilliance in the far-infrared region, the SOLEIL synchrotron radiation was used to improve the signal-to-noise ratio of the spectrum at the maximal resolution of \SI{0.001}{\wn}.

The three bands were analysed independently using two different formalisms. We used in one instance Waston’s model for symmetric-top molecules, giving common spectroscopic constants, then in another instance the tensorial formulation and group theory methods developed in the Dijon group\cite{hrs021,Wenger2008}, that allow us to provide a set of effective spectroscopic parameters. We will discuss the comparison of both models and will explain why the first overtone $2\nu_{20}$ ($A_1+E$, \SI{595}{\wn}) was measured but could not be analysed with same quality as fundamental bands.

\section{Experimental details}
Gas-phase trioxane ($>99\%$, Aldrich), was injected in a multipass cell equipped with  polypropylene films of \SI{50}{\micro\metre} thickness as cell windows, for which the White cell optics (White-type arrangement) were set to obtain a \SI{150}{\metre} long absorption path. The three low-frequency $\nu_{20}$, $\nu_{7}$ and $\nu_{19}$ fundamental bands of trioxane are predicted to have very different infrared intensities, respectively calculated to 0.1, 19 and \SI{8}{\kilo\metre/\mol}\cite{mohamed2005some}, which required us to record three spectra at different trioxane pressures to maximize absorption signal of each band: \SI{815}{\micro\bar} for the very weak $\nu_{20}$ mode, \SI{10}{\micro\bar} for both intense $\nu_{7}$ and $\nu_{19}$ modes and \SI{50}{\micro\bar} for the 2$\nu_{20}$ overtone. The three far-IR spectra have been recorded at the maximal resolution of \SI{0.001}{\wn} using the SOLEIL synchrotron FIR radiation extracted by the AILES beamline as the continuum source of the FT interferometer equipped with a He-cooled bolometer detector and a \SI{6}{\micro \metre} mylar beamsplitter, resulting in a significant improvement of the signal-to-noise ratio in comparison with a globar source\cite{brubach2010performance,smirnova2010gas}. Consequently, the acquisition times for the FTIR spectra of $\nu_{20}$, $\nu_{7}$, $\nu_{19}$ and 2$\nu_{20}$ bands of trioxane were only \SI{9}{\hour}, \SI{11}{\hour} and \SI{11}{\hour30\minute}, respectively. Both spectra were calibrated using accurate far-IR water lines absorption\cite{horneman2005transferring}. Thanks to the high signal-to-noise ratio obtained, the line position accuracy was estimated to \SI{0.0002}{\wn} for all trioxane lines observed.

\begin{figure*}[ht]
\centerline{\includegraphics[width=0.9\textwidth]{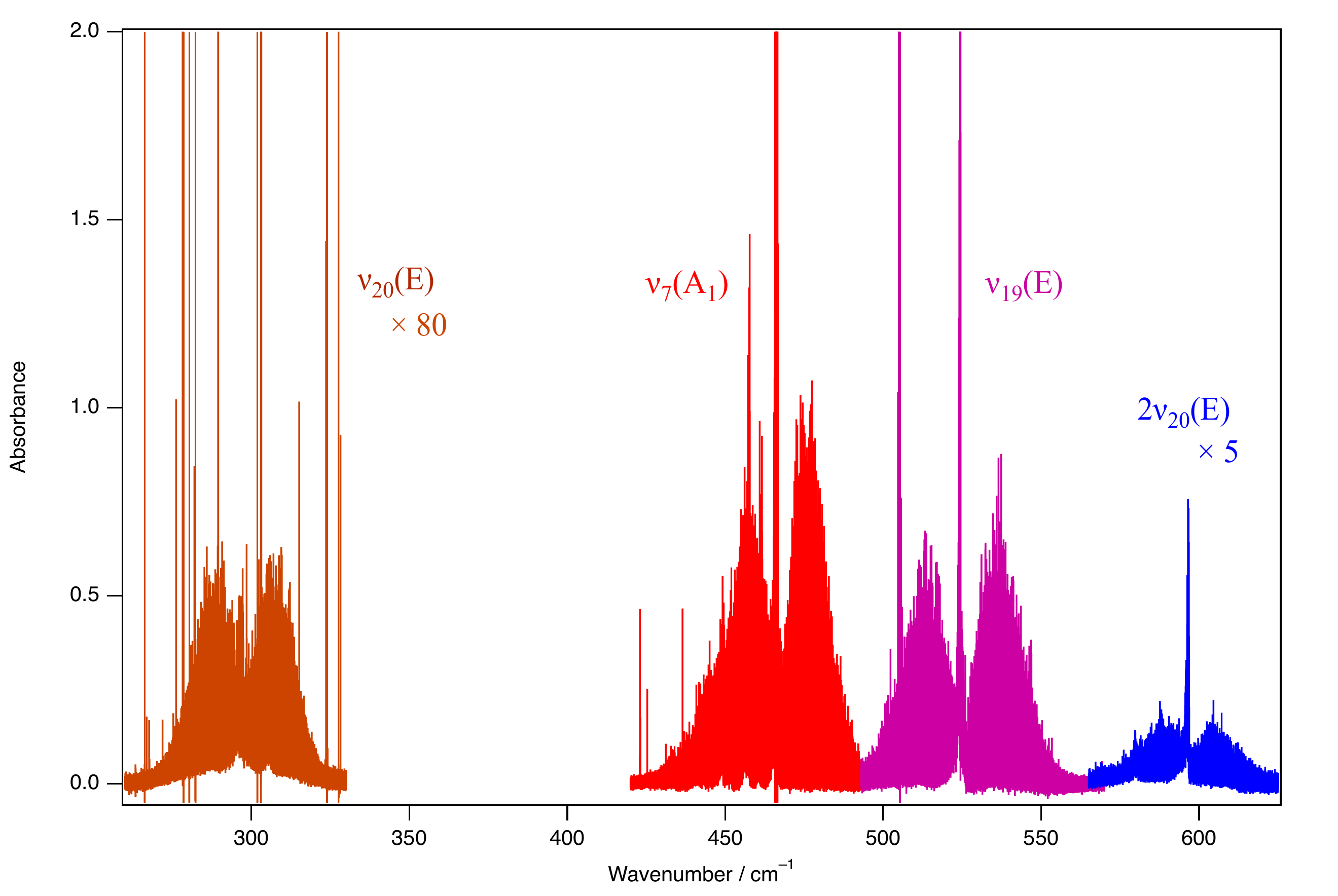}}
\caption{Overview of the high resolution spectra of the $\nu_{20}$, $\nu_7$, $\nu_{19}$ and 2$\nu_{20}$ bands of a sample of trioxane. The different bands were recorded at different pressure conditions, \SI{815}{\micro\bar} for $\nu_{20}$, \SI{10}{\micro\bar} for $\nu_{7}$ and $\nu_{19}$ and \SI{50}{\micro\bar} for 2$\nu_{20}$. The huge irregular absorption lines on the left of the spectra are water lines.}
\label{fig:trioxane_spectra}
\end{figure*}

\begin{figure*}[ht]
\centerline{\includegraphics[width=0.9\textwidth]{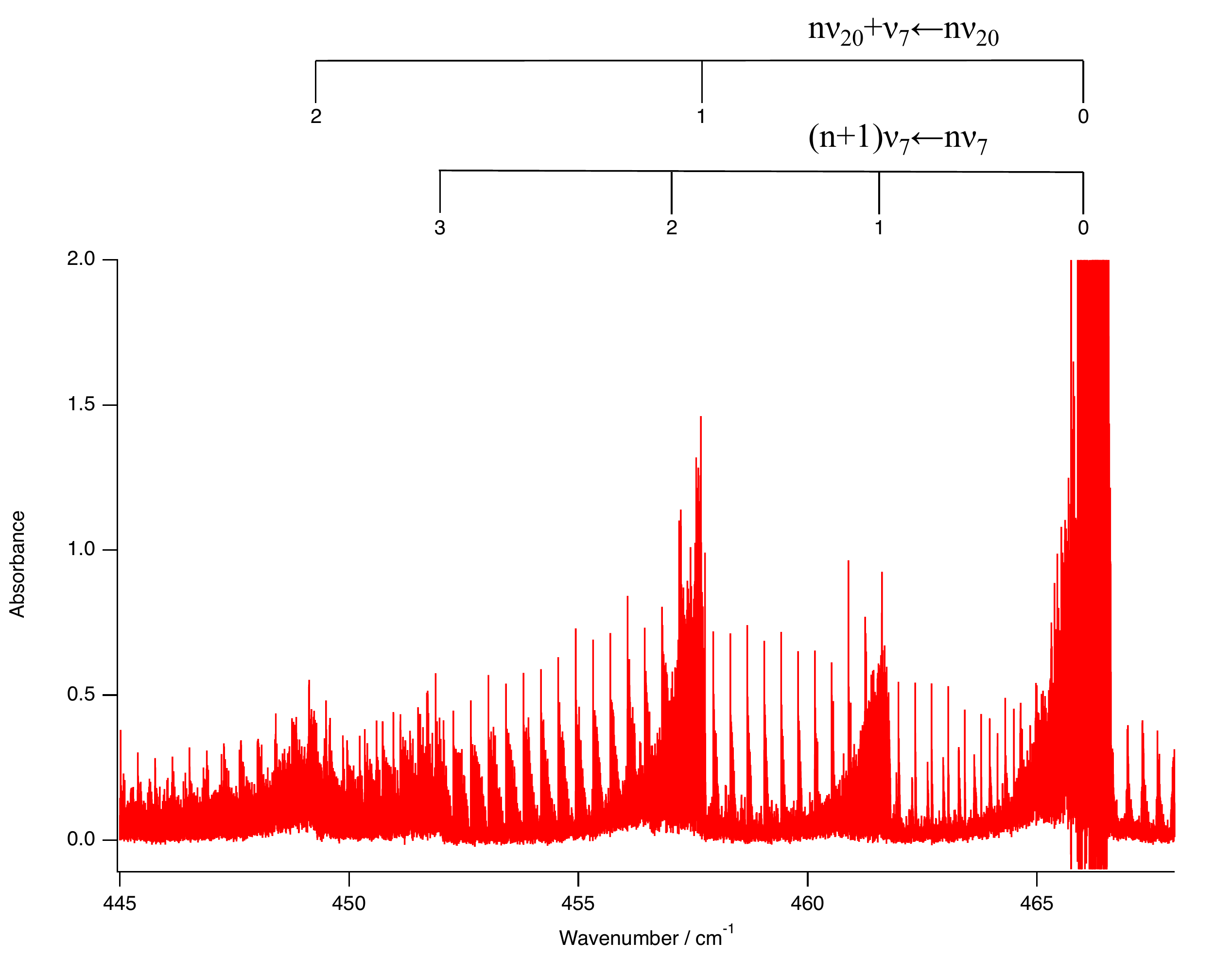}}
\caption{Two series of hot bands visible in the $P$ branch of the $\nu_7$ parallel branch. They have been tentatively assigned to $n\nu_{20} + \nu_{7} \leftarrow n\nu_{20}$ and $(n+1)\nu_{7} \leftarrow n\nu_{7}$ transitions.}
\label{fig:hot_bands}
\end{figure*}

Fig.~\ref{fig:trioxane_spectra} displays an overview of the far-IR FT spectra of trioxane recorded at high resolution. The characteristic $PQR$ band structure of three fundamentals observed is fully resolved, enabling a global rovibrational analysis. However, hot bands starting from the lowest frequency $\nu_{20}$, $\nu_{7}$ and $\nu_{19}$ modes are expected in a room temperature absorption spectrum. These features are particularly visible in the $P$ branch of the $\nu_{7}$ parallel band (Fig.~\ref{fig:hot_bands}) where two series of red shifted $Q$ branches are observed, the first one at about 461.8, 457.2 and \SI{452.0}{\wn} and tentatively assigned to $n\nu_{20} + \nu_{7} \leftarrow n\nu_{20}$ transitions up to $n=3$, the second one at 457.8 and \SI{449.3}{\wn} of the type $(n+1)\nu_{7} \leftarrow n\nu_{7}$ up to $n=2$ with anharmonicities estimated to -4.8(1) and -8.7(1)~\si{\wn}, respectively. Due to the stronger density of lines in the rotational branches of perpendicular bands, hot bands are much less visible in the $\nu_{20}$ and $\nu_{19}$ spectra but their presence could be easily evidenced by subtracting the fundamental band contour simulated from the experimental ones. Consequently, the analysis of the perpendicular bands was expected to be less straightforward than the parallel one, particularly in the $Q$ branch region due to the high density of rotational lines at high $J$ values and the additional rotational structure of hot bands.

\section{Theoretical Models}
The idea of working with two different models comes from the difficulty encountered during the analysis of $2\nu_{20}$ as explained in the section~\ref{seq:overtone_issue}. While the analysis of $\nu_{7}$, $\nu_{19}$ and $\nu_{20}$ was straightforward using Watson's formalism, the analysis of the first overtone did not yield acceptable results. Therefore, we decided to try to analyse the molecule with another model as explained below.

\subsection{Tensorial formalism}
Because of its high symmetry (a C$_{3v}$ symmetric-top), trioxane is a good candidate to be analyzed by the C$_{3v}$ Top Data System (hereafter \texttt{C$_{3v}$TDS}) software\cite{el2010c3v} developed in the Dijon group. Let us recap brieﬂy the principles of this formalism.

The theoretical model used in \texttt{C$_{3v}$TDS} is based on the tensorial
formalism and vibrational extrapolation methods described for instance in
\cite{BCGp04,CLP92,CLB99}. In the following,
\begin{itemize}
\item[$\bullet$]$\Gamma ($ \text{$=0^{+}$, $0^{-}$, $1$, $2$, $\cdots)$}
    denotes $C_{\infty v}$ irreducibles representations (irreps),
\item[$\bullet$]$C($ \text{$=A_{1}$, $A_{2}$, $E)$} is used for $C_{3v}$
    irreps.
\end{itemize}
All operators are symmetrized in the $O(3)\supset C_{\infty v} \supset C_{3v}$
group chain. The $O(3)$ standard basis set $ |J,M\rangle$ is oriented in the
subgroups through the relation
\begin{equation}
|J,\Gamma,C,\sigma\rangle =\sum_{\delta=+,-}\text{\ }^ {(\Gamma)}V_{C \sigma }^{\delta }\sum_{M=-\Gamma,+\Gamma}\text{\ }^{(J)}W_{\Gamma\delta }^{M}|J,M\rangle,
\end{equation}
where $\sigma$ is the component of $C$ if this one is degenerate (in practice, when $C$ = $E$). The $
^{(J)}W_{\Gamma\delta }^{M}$ are given by (note that there is a phase change compared to \cite{EBL05a}):
\begin{eqnarray}
\Gamma\neq 0^{\pm} & ~~~:~~~ & {}^{(J)}W^M_{\Gamma\delta}=\left|
\begin{array}{cc}
\displaystyle\frac{M}{|M|\sqrt{2}},&\delta=+,\\
\displaystyle\frac{(-1)^J}{\sqrt{2}},&\delta=-,
\end{array}
\right.\nonumber\\
\Gamma= 0^{\pm} & ~~~:~~~ & {}^{(J)}W^M_{0^{\pm}}=1.
\end{eqnarray}
The $^{(\Gamma)}V_{C\sigma }^{\delta }$ matrix elements are given in \cite{EBL05a}. The tensorial operators are oriented in the same way.

Let us consider a molecule whose vibrational levels are
grouped in a series of polyads designed by $P_{k}(k=0,...,n,...)$, $P_{0}$
being the ground-state (GS). The Hamiltonian can be developed as a
sum of operators specific to each polyad as:
\begin{equation}
\mathcal{H}=\mathcal{H}_{\{P_{0}=GS\}}+\mathcal{H}_{\{P_{1}\}}+...+\mathcal{H}_{\{P_{k}\}}+...+\mathcal{H}_{\{P_{n-1}\}}+\mathcal{H}_{\{P_{n}\}}+...
\end{equation}
We can now define an effective Hamiltonian for a given vibrational polyad (or
group of vibrational levels) by
\begin{equation}
\tilde{\mathcal{H}}^{<polyad>}=P^{<polyad>}\mathcal{H}P^{<polyad>},
\end{equation}
where
\begin{equation}
P^{<polyad>}=\displaystyle\sum_{i}|\psi_{v}^{i}\rangle\langle\psi_{v}^{i}|
\end{equation}
is the projection operator on the vibrational Hilbert subspace,
$\{|\psi_{v}^{i}\rangle\}$, for the polyad under consideration. The effective
Hamiltonian for a given polyad $P_{n}$ can be written as a sum of contributions
of the different polyads, if the contact transformation has been built to
remove inter-polyad interactions :
\begin{equation}\label{eq:effective_Hamiltonian}
\tilde{\mathcal{H}}^{<P_{n}>}=\tilde{\mathcal{H}}^{<P_{n}>}_{\{GS\}}+\tilde{\mathcal{H}}^{<P_{n}>}_{\{P_{1}\}}+\cdots+\tilde{\mathcal{H}}^{<P_{n}>}_{\{P_{n}\}}.
\end{equation}
The contribution of polyad $P_{n}$ necessarily contains the operators and
parameters of the lower polyads. The different terms are written as
\begin{equation}
\widetilde{H}=\sum_{\text{all indexes}}\widetilde{t}_{\left\{ n_{s}\right\} \left\{ m_{s}\right\}
}^{\Omega (L,\Gamma_{R} )(\Gamma _{1}\Gamma _{2}\Gamma _{V})\Gamma}T_{\left\{
n_{s}\right\} \left\{ m_{s}\right\} }^{\Omega (L,\Gamma _{R})(\Gamma
_{1}\Gamma _{2}\Gamma _{V})\Gamma}.
\end{equation}
All the indexes represent the intermediate quantum numbers and symmetries
resulting from the construction. The $\widetilde{t}$ 's are the parameters of
the model. Each $T$ operator is constructed as a tensorial coupling between a
rotational ($R$) and a vibrational ($V$) operator:
\begin{equation}\label{eq:T_operator}
T_{\left\{ n_{s}\right\} \left\{ m_{s}\right\} }^{\Omega (L,\Gamma_{R}
)(\Gamma _{1}\Gamma _{2}\Gamma _{V})\Gamma}=\beta (R^{\Omega (L,\Gamma_{R}
)}\otimes {}^{\varepsilon }V_{\left\{ n_{s}\right\} \left\{ m_{s}\right\}
}^{\Gamma _{1}\Gamma _{2}(\Gamma_{V} )})^{(\Gamma,A_{1})}.
\end{equation}

where
\begin{equation}\label{eq:beta}
\beta =\left\{
\begin{array}{ccc}
\sqrt{[\Gamma _{1}]}(-\frac{\sqrt{3}}{4})^{(\frac{\Omega }{2})\text{ \ \ }} &
\text{if} & \text{\ } L=0,  \\
1 & \text{if} & L\neq 0,
\end{array}
\right.
\end{equation}
is used to let scalar terms be equal to their equivalent in the ``usual'' non
tensorial formalism \cite{PA82}. $R^{\Omega(L,\Gamma_{R})}$ and $^{\varepsilon
}V_{\left\{ n_{s}\right\} \left\{ m_{s}\right\} }^{\Gamma _{1}\Gamma
_{2}(\Gamma_{V} )}$ are rotational and vibrational operators of respective
maximum degree $\Omega$ in the rotational angular momentum components
$J_{x}$, $J_{y}$ and $J_{z}$ and $\Omega_{v}$ degree in creation and
annihilation vibrational operators. The order of each individual term is defined as
$\Omega+\Omega_{v}-2$. Let us note that here we use a coupling scheme slightly
different from that of our paper \cite{EBL05b}. \textit{i.e.} all our couplings
are made in the $C_{\infty v}$ group, then we carry out the
$C_{\infty v}\supset C_{3v}$ reduction (in \cite{EBL05b}, all couplings were realized in $C_{3v}$).

Such a Hamiltonian development scheme enables the treatment of any polyad
system. In this work and as an example, we will use the following effective
Hamiltonians:
\begin{itemize}
\item[$\bullet$] The ground-state effective Hamiltonian
\begin{equation}
\mathcal{H}^{\langle GS\rangle}= \mathcal{H}^{\langle GS\rangle}_{\{GS\}},
\end{equation}
\item[$\bullet$] The fundamental $\nu_{i}$ band effective Hamiltonian (with $i=7, 19$ or 20 in our case)
\begin{equation}
\mathcal{H}^{\langle \nu_{i}\rangle}= \mathcal{H}^{\langle \nu_{i}\rangle}_{\{GS\}}+\mathcal{H}^{\langle \nu_{i}\rangle}_{\{\nu_{i}\}}.
\end{equation}
\end{itemize}
$H_{\{\nu_{i}\}}^{\langle \nu_{i}\rangle}$ is constructed using $^{\varepsilon
}V_{\left\{ i\right\} \left\{ i\right\} }^{\Gamma_{1}\Gamma_{2} (\Gamma_{v})}$
vibrational operators which involve creation $a^+$ and annihilation $a$ operators symmetrized in $C_{\infty v}$. The rovibrational basis functions are built according to the same coupling scheme as for the operators:
\begin{equation}
\label{base}
\left| \left[ \Psi _{r}^{(J_{r},\Gamma_{r})}\otimes \Psi
_{v}^{(\{v_{s}\},\Gamma_{v})}\right] _{\sigma }^{(\Gamma,C)}\right\rangle ,
\end{equation}

where $\Psi _{r}^{(J_{r},\Gamma_{r})}$ is the rotational basis set and $\Psi
_{v}^{(\{v_{s}\},\Gamma _{v})}$ is the vibrational one. For the twenty normal
modes of vibration of trioxane, we have:
\begin{align}
\Psi _{v}^{(\{v_{s}\},\Gamma_{v})}=\ & \Psi_{v_{1}}^{(0^{+})}\times\cdots\times\Psi_{v_{7}}^{(0^{+})}\times\Psi_{v_{8}}^{(0^{-})}\times\Psi_{v_{9}}^{(0^{-})}\times\Psi_{v_{10}}^{(0^{-})}\\
&\times\left((\cdots(\Psi_{v_{11}}^{(\Gamma_{11})}\otimes\Psi_{v_{12}}^{(\Gamma_{12})})^{\Gamma_{11,12}}\otimes\cdots)^{(\Gamma_{11\cdots 19})}\otimes\Psi _{v_{20}}^{(\Gamma_{20})}\right)^{(\Gamma_{v})}.
\end{align}
$v_{1}$,...,$v_{20}$ are vibration quantum numbers for the trioxane molecule.
$\Gamma$ (respectively $C$) is the $C_{\infty v}$ (respectively $C_{3v}$) total symmetry and
we have the reduction:
\begin{equation}
\mathcal{D}^{(\Gamma)}\supset\mathcal{D}^{(C)}.
\end{equation}
Expressions of the matrix elements of the rovibrational Hamiltonian can be
easily calculated thanks to the Wigner-Eckart theorem. All the rovibrational levels are described by $(J,C,\alpha)$
labels where $\alpha$ is a numbering index for levels that have the same $C_{3v}$
symmetry $C$ within a $J$ block. In this way, the usual $K$ quantum number is
hidden in the output and related to $\alpha$ and $C$ symmetry. So, the $K$
values do not appear explicitly in our labels and the $\triangle K$ nomenclature
does not occur in our transition labels (although $K$ is used internally as a $C_{\infty v}$ label).

\texttt{C$_{3v}$TDS} package was successfully validated in 2016 with the study of methyl iodide (CH$_3$I) on the band $\nu_{6}=1$, lying in the mid-infrared spectral region\cite{haykal2016line}.

\subsection{Watson's formalism}
Defined in 1968 by James K.G. Watson in its milestone article\cite{watson1968simplification}, in which he simplified the vibration–rotation molecular Hamiltonian, the so-called Watsonian, is now widely used for studying polyatomic molecules.

The ro-vibrational structure of the trioxane molecule was analyzed using the following Hamiltonian:
\begin{widetext}
\begin{subequations}\label{eq:HWatson}
 \begin{align}
  \hat{H}_\mathrm{Watson} =\ & T_v + BJ(J+1) + (C-B)K^2 \label{eq:HWatson:1}\\
  &-D_JJ^2(J+1)^2-DJ-D_{JK}J(J+1)K^2-D_KK^4 \label{eq:HWatson:2}\\
  &+ H_JJ^3(J+1)^3 + H_{JK}J^2(J+1)^2K^2 + H_{KJ}J(J+1)K^4 + H_KK^6 \label{eq:HWatson:3}\\
  & -2C\zeta Kl + \eta_JJ(J+1)Kl + \eta_KK^3l \label{eq:HWatson:4}\\
  & + \frac{1}{2}\left[q + D_{q_J}J(J+1)+D_{q_K}K^2\right]\left( L^2_+J^2_-+L^2_-J^2_+\right).\label{eq:HWatson:5}
 \end{align}
\end{subequations}
\end{widetext}

Terms (\ref{eq:HWatson:1}), (\ref{eq:HWatson:2}) and (\ref{eq:HWatson:3}) are standard terms for symmetric-top molecules that are used for ground- and excited states. In our work, the development to the sextic constants was needed because they have been derived in the ground state from the millimeter-wave study of Klein et \emph{al.}\cite{klein1996terahertz}. The other terms, only appropriate for excited states, represent the Coriolis parameters (\ref{eq:HWatson:4}) and $l$-doubling constants (\ref{eq:HWatson:5}). Nevertheless, these are only needed for perpendicular bands. We purposely omitted here unused terms, like pure rotational $C_{3v}$ splitting tems $\varepsilon, \varepsilon_J, \ldots$, that were not fitted in the present study.

\subsection{Conversion between both formalisms}\label{seq:conversion}
Although the two models are different, they are both built on the same general principles.  It is thus possible to convert spectroscopic parameters from one formalism to another. This is done by expanding both the Watsonian and our tensorial Hamiltonian in terms of elementary rotational operators $J^+$, $J^-$ and $J_z$ and then by identifying the coefficients of each term. In practice, this is performed thanks to the Maxima symbolic computation software\cite{maxima}. In this section, we will list the resulting formulas needed to convert tensorial parameters to Watsonian parameters (and vice versa) in the case of the trioxane molecule. Further details on how these formulas were determined are described in another publication\cite{boudon2022analytical} in the case of purely rotational operators. Coriolis ($\zeta$, $\eta_J$) and $l$-doubling constants ($q_+$) are derived by comparing the eigenvalues of the corresponding operators in both models. $\eta_K$ is not so straightforward because parameter $t^{3(3,S^-)}_{i}$ is a more complex combination of Coriolis and higher order $l$-doubling constants like $t^{4(2,D)}_i$, so its expressions requires a complete analytical development of ro-vibrational operator that will be attempted in the near future.

\subsubsection{Tensorial to Watson}\label{sec:tens_to_watson}
The Watson's parameters used in our fit (except $\eta_K$, see above), in terms of tensorial parameters:
\begin{equation}
\begin{dcases}
B&=-{{3\,2^{{{7}\over{2}}}\,t^{4(4,S^+)}_i}\over{\sqrt{35
 }}}-{{2^{{{3}\over{2}}}\,t^{2(2,S^+)}_i}\over{\sqrt{3}}}+t^{2(0,S^+)}_i,\\
 C&={{19\,2^{{{5}\over{2}}}\,t^{4(4,S^+)}_i}\over{\sqrt{35}}}+{{2^{{{5}\over{2}}}\,t^{2(2,S^+)}_i}\over{\sqrt{3}}}+t^{2(0,S^+)}_i, \\
 D_{J}&=-{{32\,\sqrt{6}\,
 t^{6(4,S^+)}_i}\over{\sqrt{35}}}-{{3\,2^{{{5}\over{2}}}\,t^{4(4,S^+)}_i}\over{\sqrt{35
 }}}-{{2^{{{7}\over{2}}}\,t^{4(2,S^+)}_i}\over{3}}-t^{4(0,S^+)}_i, \\
 D_{{\it JK}}&={{8\,10^{{{3}\over{2}}}\,t^{6(4,S^+)}_i}\over{
 \sqrt{21}}}+{{24\,\sqrt{10}\,t^{4(4,S^+)}_i}\over{\sqrt{7}}}+2^{{{7}\over{2}}
 }\,t^{4(2,S^+)}_i, \\
 D_{K}&=-4\,\sqrt{70}\,t^{4(4,S^+)}_i, \\
 \zeta&=-\frac{t^{1(1,S^-)}_i}{C\sqrt{2}}, \\
 \eta_J&= 4\sqrt{\frac{2}{3}}t^{3(1,S^-)}_i, \\
 q_+ &=2t^{2(2,D)}_i, \\
 H_{J}&=-{{16\,\sqrt{6}\,t^{6(4,S^+)}_i}\over{\sqrt{35}}}-{{2^{{{11}\over{2}}}\,t^{6(2,S^+)}_i}\over{3^{{{3}\over{2}}}}}+t^{6(0,S^+)}_i, \\
 H_{{\it JK}}&={{32\,\sqrt{30}\,t^{6(4,S^+)}_i}\over{\sqrt{7}}}+{{2^{{{11}\over{2}}}\,t^{6(2,S^+)}_i
 }\over{\sqrt{3}}}, \\
 H_{{\it KJ}}&=-{{16\,\sqrt{70}\,t^{6(4,S^+)}_i}\over{\sqrt{3}}}.
 \end{dcases}
\end{equation}

\subsubsection{Watson to tensorial}\label{sec:watson_to_tensorial}
Expression of tensorial parameters used in our fit, in terms of Watson's parameters, without $t^{3(3,S^-)}_i$ and $t^{4(2,D)}_i$ (see the explanation at the beginning of Sect.~\ref{seq:conversion}) are:
\begin{equation}\label{eq:t}
\begin{dcases} 
 t^{1(1,S^-)}_i&=-\zeta C \sqrt{2}, \\
 t^{2(0,S^+)}_i&={{D_{K}}\over{15}}+{{2\,B}\over{3}}+{{C}\over{3}}, \\
 t^{2(2,S^+)}_i&={{5\,D_{K}}\over{14\,\sqrt{6}}}-{{B}\over{2\,\sqrt{6}}}+{{C}\over{2\,\sqrt{6}}}, \\
 t^{2(2,D)}_i&=\frac{q_+}{2}, \\
 t^{3(1,S^-)}_i&=\eta_J \frac{\sqrt{3}}{4\sqrt{2}}, \\
 t^{4(0,S^+)}_i&=-{{H_{{\it KJ}}}\over{15}}-{{D_{K}}\over{5}}-{{D_{{\it JK}}}\over{3}}-D_{J}, \\
 t^{4(2,S^+)}_i&={{5\,H_{
 {\it KJ}}}\over{7\,2^{{{7}\over{2}}}}}+{{3\,D_{K}}\over{7\,2^{{{5}\over{2}}}}}+{{D_{{\it JK}}}\over{2^{{{7
 }\over{2}}}}}, \\
 t^{4(4,S^+)}_i&=-{{D_{K}}\over{4\, \sqrt{70}}},\\
 t^{6(0,S^+)}_i&={{H_{{\it KJ}}}\over{5}}+{{H_{{\it JK}}}\over{3}}+H_{J}, \\
 t^{6(2,S^+)}_i&={{3^{{{3}\over{2
 }}}\,H_{{\it KJ}}}\over{7\,2^{{{9}\over{2}}}}}+{{\sqrt{3}\,H_{{\it JK}}}\over{2^{{{11}\over{2
 }}}}}, \\
 t^{6(4,S^+)}_i&=-{{\sqrt{3}\,H_{{\it KJ}}}\over{16\,\sqrt{70}}}.
\end{dcases}
\end{equation}

The Dijon tensorial model relies on the so-called ``vibrational extrapolation’’ (see Eq.~\ref{eq:effective_Hamiltonian}) that possesses a clear advantage to ensure the convergence of fits and consistency of parameters. In practice, this means that effective Hamiltonian parameters for a given vibrational state or polyads are (presumably small) corrections to the parameters of the states (or polyads) below. Just as a simple illustration, we can say that the traditional approach consists in fitting parameters for each state: $B_0, D_0, \ldots, B_1, D_1, \ldots$. The vibrational extrapolation methods rather consists in fitting $B_0, D_0, \ldots, \Delta B_1=B_1-B_0, \Delta D_1=D_1-D_0, \ldots$. Ensuring that parameter values for each state are not too far from the ground state ones amounts to verify that $\Delta B_1=B_1-B_0, \Delta D_1=D_1-D_0, \ldots$ are small. Fixing these difference to zero, like $\Delta D_1=0$ for instance forces excited state parameters to be identical to the ground state ones in the fit when such a constraint appears necessary.

This principle is applied in the present study. Thus, for the connection with Watson’s formalism, in order to apply the conversion formulas, we first started to convert our ``difference’’ excited state parameters, let us call them $\Delta t_i^k$ ($i=7$, 19 or 20, and $k$ the parameter index representing $\Omega(L,\Gamma_R)$) to an ``absolute’’ excited state value by adding the ground state value ($t_i^k=\Delta t_i^k+t_0^k$, were $t_0^k$ is a ground-state parameter). Then, the conversion formulas give Watsonian parameters for state $i$, like B$_i$, {\em etc\/}. Of course, some excited state parameters like Coriolis ones have no counterpart in the ground state and are thus not differences but directly ``absolute’’ excited state values.

In the case of fundamental excited states ($v=1$) with $E$ symmetry ($\nu_{19}$ and $\nu_{20}$ in our case), all $\Delta t_i^k$ have to be multiplied by the matrix element of the vibrational operator, which is $1/\sqrt{2}$ and, when $L=0$, one also has to take into account the $\sqrt{[\Gamma_1]}=\sqrt{2}$ factor which is present in the $\beta$ factor (see Eqs.~(\ref{eq:T_operator}) and (\ref{eq:beta}) above). The general conversion formulas for any vibrational state will be the subject of a future paper.

\section{Analysis and discussion}
Spectra analyses were conducted with different software packages. First, for the tensorial formalism, we used the software suite developed in the Dijon group ; \texttt{SPVIEW} (Spectrum-View)\cite{Wenger2008}, in its version 2.0\footnote{\url{https://icb.u-bourgogne.fr/spview}}, for the line assignment and \texttt{XTDS} (eXtended spherical-Top Data System)\cite{Wenger2008}, using \texttt{C$_{3v}$TDS} package, for spectra modeling and job executions. Then, for Watson's formalism, we used the well known \texttt{PGOPHER}\cite{western2017pgopher}, program for rotational, vibrational and electronic spectra developed by the late Dr Colin Western and \texttt{SPFIT}\cite{Pic91}, the program of H.M. Pickett originally written to fit and predict spectra of asymmetric-top molecules involving spin- and rotation-vibration interaction and treating symmetric-top ones as special cases.

\begin{figure*}[ht]
\centerline{\includegraphics[width=0.9\textwidth]{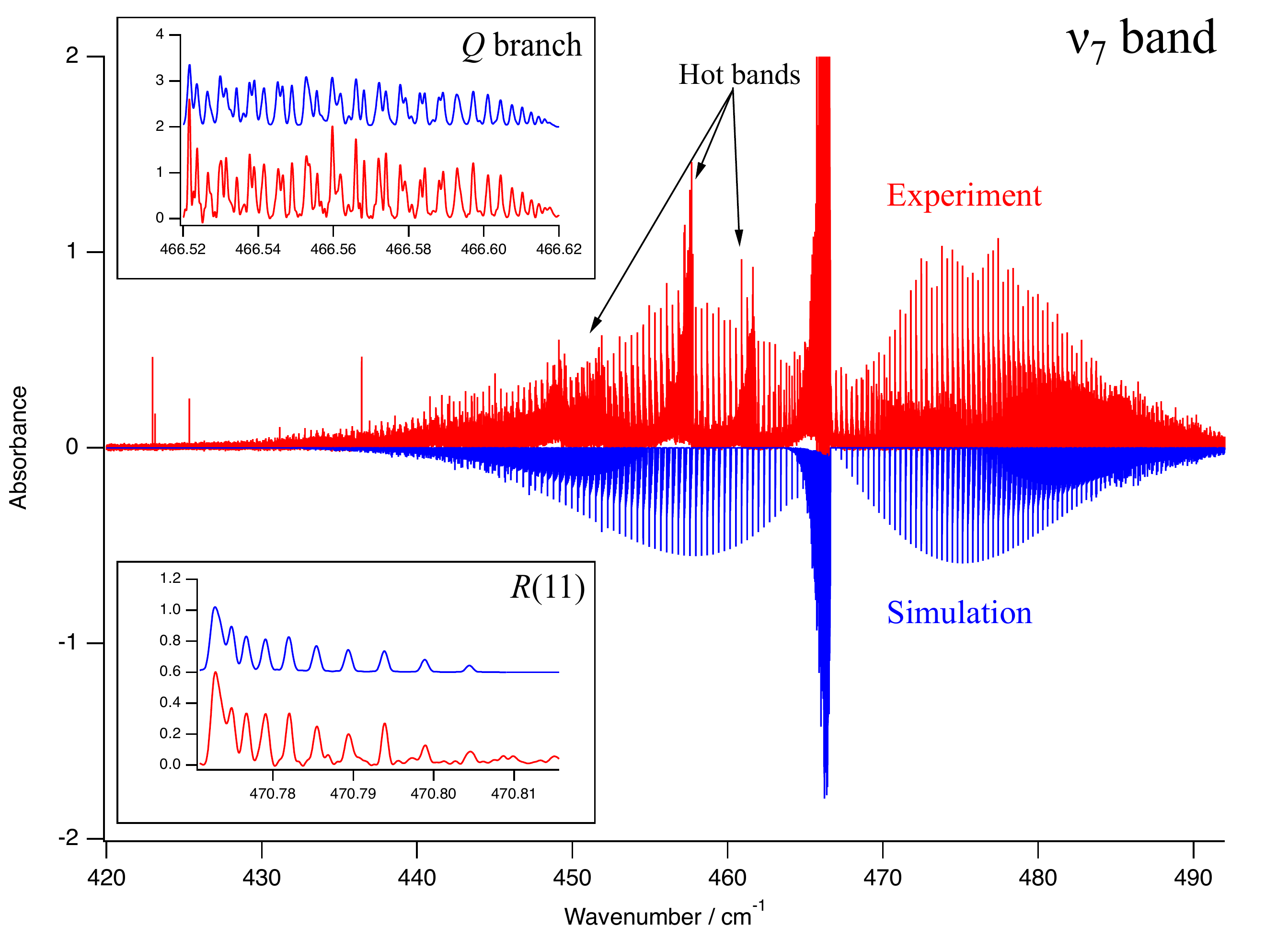}}
\caption{The $\nu_7$ band, compared to the simulation. The insets display a few lines in the $Q$ branches and the $K$ series in $R(11)$.}
\label{fig:trioxane_comparison}
\end{figure*}

Fig.~\ref{fig:trioxane_comparison} compares experiment and simulation for the $\nu_7$ band. Some insets show again that the fine structure is very well modeled.

\subsection{Analysis in the tensorial formalism}\label{sec:analyse_tensorial}
This analysis takes advantage of the millimeter-wave study of Klein et~\emph{al.}\cite{klein1996terahertz} in the ground-state region around 350--\SI{950}{\giga\hertz}. Data from the paper were converted into a 2-columns flat file in order to produce a stick spectrum readable by \texttt{SPVIEW}. Then, we proceeded to a line assignment to perform a standard iterative Levenberg-Marquardt non-linear least squares fit. A total of 289 lines (218 different frequencies) were used, giving a root mean square deviation of \SI{0.015}{\mega\hertz} with eight parameters up to the sixth order (six free and two fixed), as shown in Table~\ref{tab:tensorial-global}. As pure rotational data for only one isotopologue do not allow us to fit $C$ and $D_K$, the molecule's structure remains undetermined. These parameters were derived from a study of Colmont\cite{colmont1974etude} in 1974 thanks to $^{13}$C and $^{18}$O isotopic substitutions, which break the molecular. On our side, all ground-state parameters were used in the formulas~(\ref{eq:t}) as a starting point of our fit and the two spectroscopic parameters $t^{2(2,S^+)}_i$ and $t^{4(2,S^+)}_i$ were fixed to the calculated value.

The $\nu_7$, $\nu_{19}$ and $\nu_{20}$ bands were analysed successively, first by manually estimating the band center and using rotation constants from ground-state in order to produce a fit and a first prediction. Then, the fit was iteratively improved, releasing more parameters, by adding new identified lines.

\begin{figure*}[ht]
\centerline{\includegraphics[width=0.9\textwidth]{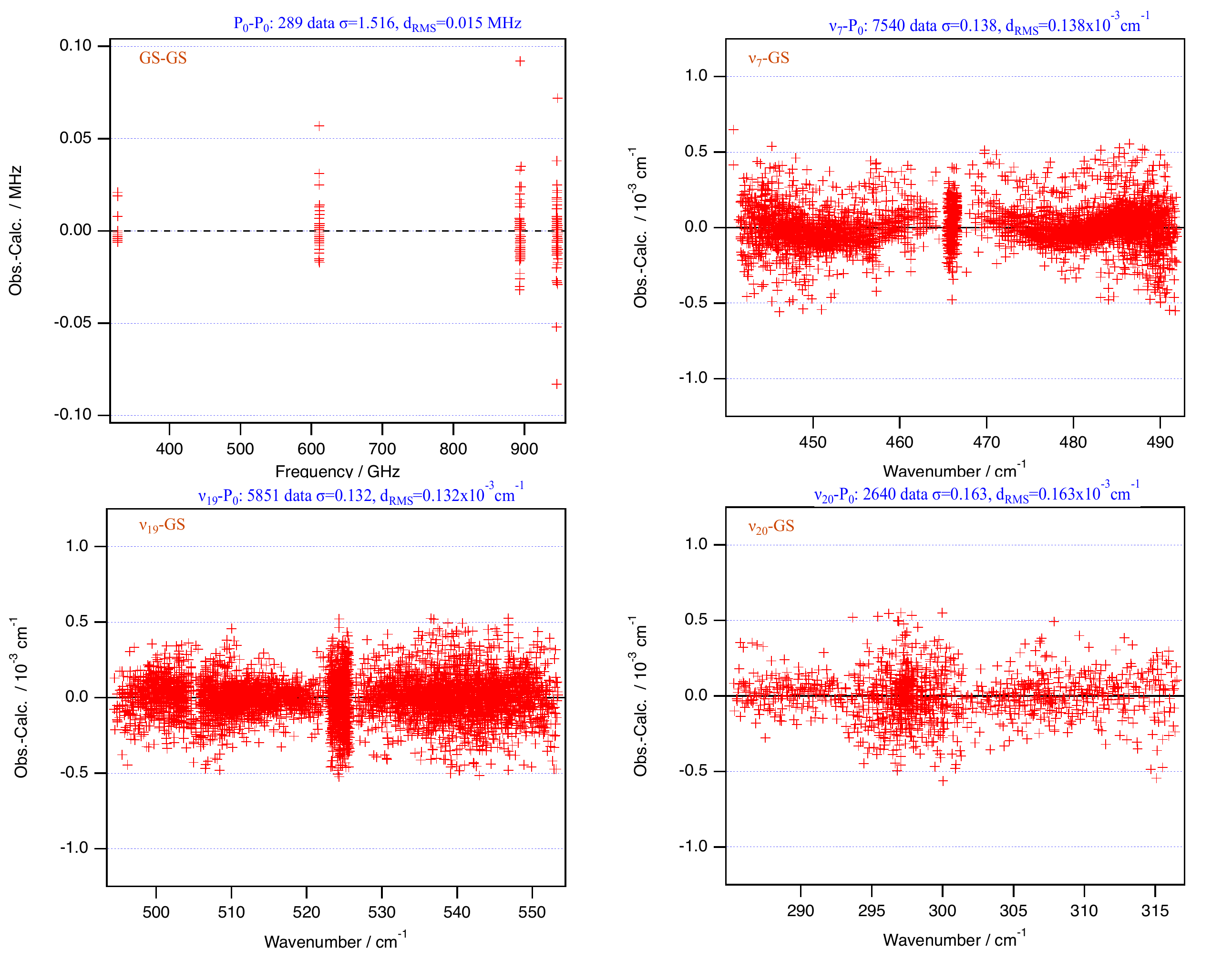}}
\caption{Fit residuals for line positions for the rotational and the three different types of ro-vibrational transitions used in the present work.}
\label{fig:residuals_tensorial}
\end{figure*}

Then, we performed a global fit using a dedicated polyad scheme where $P_0$ is the ground-state, $P_3-P_0$ stands for the $\nu_{20}$ band, $P_4-P_0$ for $\nu_{7}$ and $P_5-P_0$ for $\nu_{19}$. In this scheme it is possible to add the first overtone of $\nu_{20}$ in $P_6-P_0$. We obtained an excellent ﬁt whose root mean squares deviation are \SI{0.015}{\mega\hertz}, 1.32$\times 10^{-4}$\si{\wn}, 1.38$\times 10^{-4}$\si{\wn} and 1.63$\times 10^{-4}$\si{\wn} for the transitions GS$-$GS, $\nu_{7}-$GS, $\nu_{19}-$GS and $\nu_{20}-$GS, respectively. The global standard deviation is of 0.245 for a total of 16\,320 lines. Fig.~\ref{fig:residuals_tensorial} details the fit residuals for line positions for these
four types of transitions, along with some statistics, then all the fit results are gathered in Table~\ref{tab:tensorial-global}. 

Finally, $t^{4(2,D)}_i$, $t^{4(4,D)}_i$ and $t^{5(1,S^-)}_i$ are fitted in the $\nu_{19}$ band, but not in the $\nu_{20}$. Indeed, adding these parameters to the latter does not improve the fit and they are not well defined.  In the same way, removing them from the $\nu_{19}$ one degrades the fit by a factor 11.

\begin{landscape}
\renewcommand{\arraystretch}{1.25}
\begin{table}[ht]
 \caption{Spectroscopic constants of the ground-vibrational state, $\nu_7$, $\nu_{19}$ and $\nu_{20}$ bands of Trioxane fitted in the tensorial formalism.}
\label{tab:tensorial-global}
\begin{center}
\resizebox{\columnwidth}{!}{\footnotesize\begin{tabular}{
	l
	r
	l
	r
	l
	r
	l
	r
	l
	}
\hline\hline
Parameters & \multicolumn{8}{c}{Value $/$ cm$^{-1}$ (Hamiltonian $\widetilde{H}$)}  \\\cline{2-9}
\multicolumn{1}{c}{$t^{\Omega{( K, nC )}}_i$}& \multicolumn{2}{c}{GS} & \multicolumn{2}{c}{$\nu_7$} & \multicolumn{2}{c}{$\nu_{19}$} & \multicolumn{2}{c}{$\nu_{20}$}   \\
\hline
 $t^{0(0,S^+)}_i$  &                    &                    & 466.618966(37) &                    & 524.466369(30) &                    & 297.703777(50) &                    \\
 $t^{1(1,S^-)}_i$  &                    &                    &                &                    & 9.25909(12)    & $\times$10$^{ -2}$ &    3.91660(11) & $\times$10$^{ -2}$ \\
 $t^{2(0,S^+)}_i$  & 1.498526396(12)    & $\times$10$^{ -1}$ &-3.46744(31)    & $\times$10$^{ -4}$ & 4.7671(27)     & $\times$10$^{ -5}$ &   -1.38378(70) & $\times$10$^{ -4}$ \\
 $t^{2(2,S^+)}_i$  &-1.5948802396$^\dag$ & $\times$10$^{ -2}$ & 5.3640(12)     & $\times$10$^{ -5}$ & -6.4955(13)    & $\times$10$^{ -5}$ &     2.7883(31) & $\times$10$^{ -5}$ \\
 $t^{2(2,D)}_i$  &                    &                    &                &                    & -2.16888(36)   & $\times$10$^{ -4}$ &   -1.29879(74) & $\times$10$^{ -4}$ \\
 $t^{3(1,S^-)}_i$  &                    &                    &                &                    & 4.060(27)      & $\times$10$^{ -8}$ &      2.288(15) & $\times$10$^{ -8}$ \\
 $t^{3(3,S^-)}_i$  &                    &                    &                &                    & -1.018(14)     & $\times$10$^{ -8}$ &      -8.65(12) & $\times$10$^{ -9}$ \\
 $t^{4(0,S^+)}_i$  &-2.357311(25)       & $\times$10$^{ -8}$ & 1.2388(57)     & $\times$10$^{ -9}$ & -7.470(53)     & $\times$10$^{-10}$ &       1.32(23) & $\times$10$^{-10}$ \\
 $t^{4(2,S^+)}_i$  &-5.5033842448$^\dag$ & $\times$10$^{ -9}$ & 2.262(12)      & $\times$10$^{-10}$ & -1.454(14)     & $\times$10$^{-10}$ &       4.23(60) & $\times$10$^{-11}$ \\
 $t^{4(2,D)}_i$  &                    &                    &                &                    & -5.064(40)     & $\times$10$^{-10}$ &                   & \\
 $t^{4(4,S^+)}_i$  &-1.74069(41)        & $\times$10$^{-10}$ & 6.138(47)      & $\times$10$^{-11}$ & -4.658(52)     & $\times$10$^{-11}$ &     1.84(19) & $\times$10$^{-11}$   \\
 $t^{4(4,D)}_i$  &                    &                    &                &                    & -1.016(52)     & $\times$10$^{-10}$ &                &                    \\
 $t^{5(1,S^-)}_i$  &                    &                    &                &                    & -1.38(12)      & $\times$10$^{-13}$ &                &                    \\
 $t^{6(0,S^+)}_i$  & 1.1285(18)         & $\times$10$^{-14}$ &                &                    &                &                    &                &                    \\
 $t^{6(2,S^+)}_i$  & 3.146(20)          & $\times$10$^{-16}$ &                &                    &                &                    &                &                    \\
 $t^{6(4,S^+)}_i$  &-1.17956(68)        & $\times$10$^{-15}$ &                &                    &                &                    &                &                    \\
 \\
 Lines fitted & \multicolumn{2}{c}{289} & \multicolumn{2}{c}{5851} & \multicolumn{2}{c}{7540} & \multicolumn{2}{c}{2640}   \\
 $J_\mathrm{max}$ & \multicolumn{2}{c}{89} & \multicolumn{2}{c}{80} & \multicolumn{2}{c}{90} & \multicolumn{2}{c}{88} \\
 free parameters & \multicolumn{2}{c}{6} & \multicolumn{2}{c}{6} & \multicolumn{2}{c}{13} & \multicolumn{2}{c}{10}   \\
 $d_{\mathrm{RMS}}{}^*$ & \multicolumn{2}{c}{0.015} & \multicolumn{2}{c}{0.132} & \multicolumn{2}{c}{0.138} & \multicolumn{2}{c}{0.163} \\
 \cline{1-3}
 Total of Lines & 16\,320 \\
 Standard deviation & 0.245 \\
\hline
\multicolumn{8}{l}{\footnotesize{$^*$ $d_{\mathrm{RMS}}$ is given in $10^{-3}$\si{\wn} except GS that is in MHz.}}\\
\multicolumn{8}{l}{\footnotesize{$^\dag$ Fixed value.}}
\end{tabular}}

\end{center}

\end{table}
\end{landscape}

\subsection{Analysis in the Watson formalism}\label{sec:watsonian_formalism}
Preliminary simulations of the three far-IR bands of trioxane were carried out separately using \texttt{PGOPHER} program configured with the Watson’s Hamiltonian dedicated to oblate top-symmetric molecules. The rovibrational analysis of the $\nu_{7}$ OCO bending mode ($A$ symmetry) centered at about \SI{466}{\wn} is realized on grounds of a parallel band using only (\ref{eq:HWatson:1}), (\ref{eq:HWatson:2}) and (\ref{eq:HWatson:3}) standard terms of $\hat{H}_\mathrm{Watson}$. For both $\nu_{20}$ CH$_2$ torsion and $\nu_{19}$ OCO bending modes ($E$ symmetry) centered at about 297 and \SI{524}{\wn}, the presence of Coriolis parameters and $l$-doubled states requires to use (\ref{eq:HWatson:4}) and (\ref{eq:HWatson:5}) non standard terms. The fit of each band is realized by fixing the ground state constants to Klein et al values and excited state sextic constants to ground state ones. At low values of $J$, $K$, only the band center and two rotational constants are adjusted for a parallel band to which the $\zeta$ Coriolis and $q_+$ $l$-doubling parameter are added for a perpendicular band. At higher values of $J$, $K$, quartic constants and higher order of $\zeta$ and $q_+$ terms are introduced. 

The three linelists obtained from the \texttt{PGOPHER} simulations are then converted in a single \texttt{.lin} file to realize a global adjustment of $\nu_{7}$, $\nu_{19}$ and $\nu_{20}$, bands with the \texttt{SPFIT} program. For this last step, the 213~millimeter-wave lines of Klein et \emph{al.} are added to the excited state line transitions so that ground- and excited-state parameters are together adjusted.

\begin{figure*}[ht]
\centerline{\includegraphics[width=0.9\textwidth]{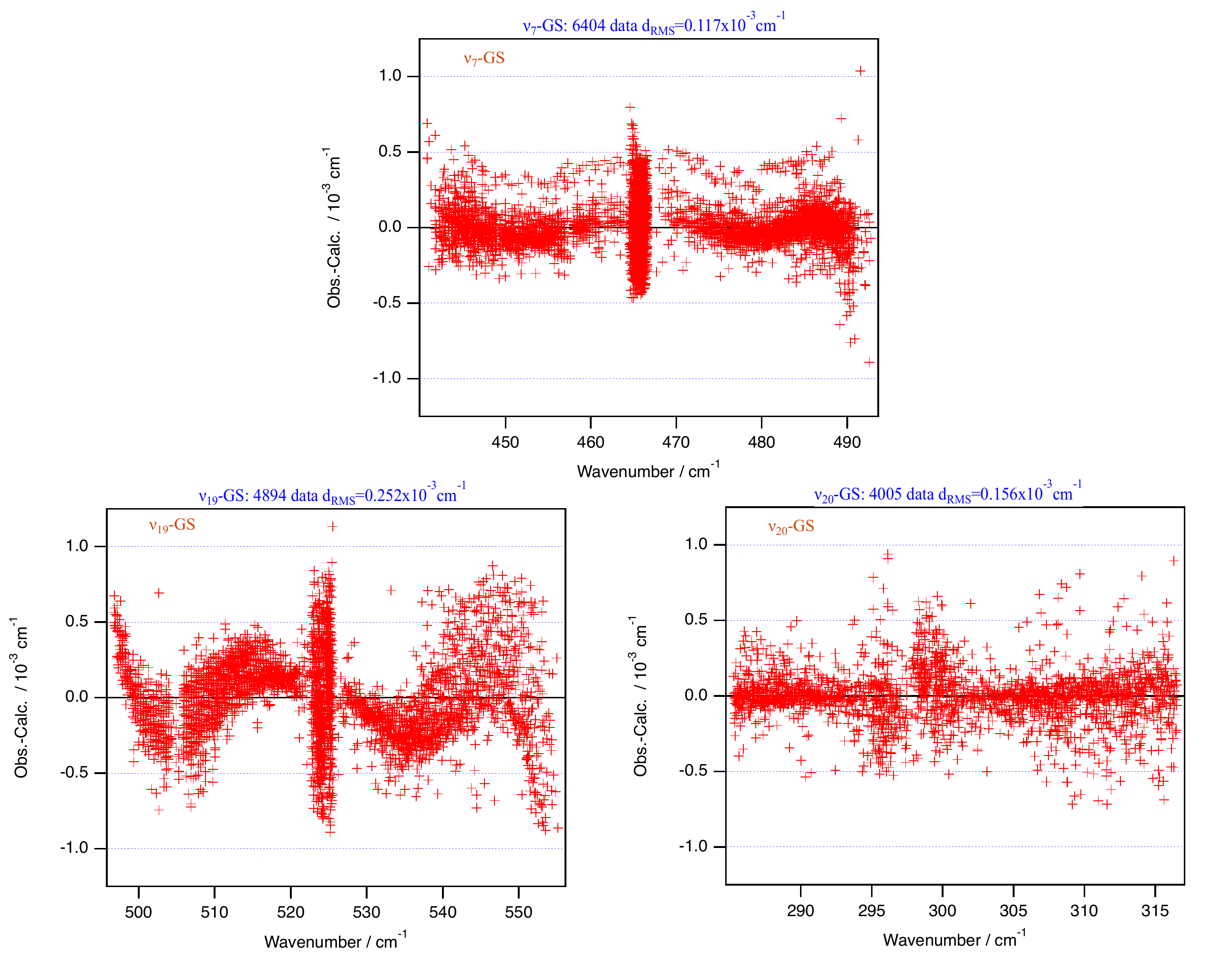}}
\caption{Fit residuals for line positions for the three different types of ro-vibrational transitions used in the present work, with the Watson's formalism.}
\label{fig:residuals_watson}
\end{figure*}

We obtained a root mean squares deviation of 1.17$\times 10^{-4}$~\si{\wn}, 2.52$\times 10^{-4}$~\si{\wn} and 1.56$\times 10^{-4}$~\si{\wn} for the transitions $\nu_{7}$--GS, $\nu_{19}$--GS and $\nu_{20}$--GS, respectively. The global standard deviation is of 0.609 for a total of 15\,512 lines. Fig.~\ref{fig:residuals_watson} details the fit residuals for line positions for these 3 transitions, along with some statistics while the spectroscopic parameters are gathered in Table~\ref{tab:watson_param}.

\begin{landscape}
\renewcommand{\arraystretch}{1.25}
\begin{table}[ht]
 \caption{Spectroscopic constants of the ground-vibrational state, $\nu_7$, $\nu_{19}$ and $\nu_{20}$ bands of Trioxane. These parameters were determined in the Watsonian formalism using \texttt{SPFIT} program.}
\label{tab:watson_param}
\begin{center}
\resizebox{\columnwidth}{!}{\begin{tabular}{
	l
	S
	S
	S
	S
	}
\hline\hline
\multicolumn{1}{c}{Parameters} & \multicolumn{1}{c}{\text{GS}} & \multicolumn{1}{c}{\text{$\nu_7$}}  & \multicolumn{1}{c}{\text{$\nu_{19}$}} & \multicolumn{1}{c}{\text{$\nu_{20}$}} \\
\hline
$T_v$ (cm$^{-1}$) &               & 466.618962(12) & 524.466456(15) & 297.703913(16) \\
$B$ (MHz)     & 5273.257177(41)   & 5260.23581(41) & 5276.94622(73) & 5268.15003(82) \\
$C$ (MHz)     & 2933.95(27)       & 2928.81(27) & 2930.89(27) & 2931.73(27) \\
$D_J$ (kHz)   & 1.3438897(92)     & 1.27586(10)     & 1.39220(28)    & 1.34295(28) \\
$D_{JK}$ (kHz)& -2.016344(20)     & -1.88696(23)   & -2.10556(59)     & -2.01170(60) \\
$D_K$ (kHz)   & 0.17$^\ddag$          & 0.10883(16)    & 0.21842(33)      & 0.16627(35) \\
$\zeta$       &                      &                & -0.66869(15)   & -0.28215(10) \\
$\eta_{J}$ (kHz)&                    &                & -2.311(8)      & -0.922(10) \\
$\eta_{K}$ (kHz)&                    &                & -3.037(8)      & -2.249(10) \\
$q_+$ (MHz)   &                      &                &  12.9115(12)   &  7.6313(31)       \\
$H_J$ (mHz)   & 0.49138(60)       & 0.49138$^\dag$ & 0.49138$^\dag$ & 0.49138$^\dag$ \\
$H_{JK}$ (mHz)& -2.0993(14)       &-2.0993$^\dag$  &-2.0993$^\dag$  &-2.0993$^\dag$ \\
$H_{KJ}$ (mHz)& 2.7352(19)        & 2.7352$^\dag$  & 2.7352$^\dag$  & 2.7352$^\dag$ \\
\\
 Lines fitted & 212                & 6405                 & 4894     & 4005 \\
 $J_\mathrm{max}$ & 89             & 80                   & 99       & 91 \\
 free parameters & 7             & 6                   & 10       & 10 \\
 $d_{\mathrm{RMS}}{}^*$ &  0.011 & 0.117                & 0.252    & 0.156 \\
 \cline{1-2}
 Total of Lines & 15512 \\
 Standard deviation & 0.609 \\
\hline
\multicolumn{5}{l}{\footnotesize{$^*$ $d_{\mathrm{RMS}}$ is given in $10^{-3}$\si{\wn} except GS that is in MHz.}}\\
\multicolumn{5}{l}{\footnotesize{$^\ddag$ Ground-state parameters fixed to the value of Klein \emph{et al.} \cite{klein1996terahertz}}.}\\
\multicolumn{5}{l}{\footnotesize{$^\dag$ $\nu_7$, $\nu_{19}$ and $\nu_{20}$ sextic parameters are fixed to the ground-state values.}.}
\end{tabular}}

\end{center}
\end{table}
\end{landscape}

\subsection{Comparison of both formalisms}

Thanks to the formulas transcribed in Sect.~\ref{sec:tens_to_watson}, we report the Watsonian spectroscopic constants converted from the tensorial ones in Table~\ref{tab:tensor-to-watson}. This allows an easier comparison between the two models. In this section, we will detail each parameter and examine the differences we observe.

First of all, not all spectroscopic constants have been converted because some of them have no equivalence in the fit we have provided in Sect.~\ref{sec:watsonian_formalism}. It is the case for $t^{3(3,S^-)}_{i}$ as already explained in Sect.~\ref{seq:conversion} and for $t^{5(1, S^-)}_{i}$ which is a Coriolis parameter at higher order than $\eta_J$. Then, the two parameters $t^{4(2, D)}_{i}$ and $t^{4(4, D)}_{i}$ represent the centrifugal distortion of $q_+$ $l$~doubling constant, called $D_{q_J}$ and $D_{q_K}$ in Eq.~(\ref{eq:HWatson}) but not fitted in this study with the Watson's model.

Second, sextic parameters are all together fixed to the values fitted for the ground-state. This is simply because no sextic parameters were fitted for the excited states in our tensorial model, as stated in Sect.~\ref{sec:watson_to_tensorial}. Regarding the residuals in Fig.~\ref{fig:residuals_watson}, $\nu_{19}-\mathrm{GS}$ is less good than the ones obtained with the other formalism in Fig.~\ref{fig:residuals_tensorial} and a polynomial shape is visible in the obs.$-$calc. plot. The fit could be improved by releasing sextic parameters, however the constants thus derived have values that look spurious, and far from those of the GS. We therefore decided to keep the results consistent for a better comparison of the models.

About rotational constants, the comparison between both formalisms of two $K-$dependent terms,  namely $(\Delta C - \Delta B)K^2$ and $-2C\zeta K$ shows a fair agreement for the parallel band $\nu_7$ (only the term in $K^2$ exists) with $(\Delta C - \Delta B)K^2$ equal to of 7.8781 and \SI{7.8788}{\mega\hertz} for TDS and Watson formalisms, respectively. For the perpendicular bands $\nu_{19}$ and $\nu_{20}$, we have illustrated in Table~\ref{tab:deltaCdeltaB} the slight differences between the $K-$dependent terms for different values of $K$.  The sum of two terms differs by about 60 and \SI{500}{\mega\hertz} between TDS and Watson formalisms for $K$ equal to 10 and 80, respectively. The term $-2C\zeta K$ is mainly responsible for this variation as indicated on Table~\ref{tab:deltaCdeltaB}. Knowing that the agreement between both formalisms is excellent for the Coriolis parameter (only 0.3\% variation for both bands), the main difference may come only from the value of $C$ in the ground state: in the Watsonian fit, $C$ is fixed to \SI{2933.95}{\wn}\cite{colmont1974etude} while in the tensorial fit, the iterative fit detailed in Sect.~\ref{sec:analyse_tensorial} imposes that $C$ in the ground state is adjusted as all other ground and excited state parameters. The difference of about \SI{3}{\mega\hertz} between both values of $C$ leads to a difference of about 300 and \SI{150}{\mega\hertz} at $K=80$ for $-2C\zeta K$, respectively for $\nu_{19}$ and $\nu_{20}$.

\renewcommand{\arraystretch}{1.25}
\begin{table}[ht]
\caption{Comparison between rotational constants for $\nu_{19}$ and $\nu_{20}$ and formalisms. For each $K$ the first line gives the value of the expression $(\Delta C - \Delta B)K^2$, while the second one is the result of $-2C\zeta K$. Values are given in \si{\mega \hertz}.}
\label{tab:deltaCdeltaB}
\begin{center}
\resizebox{\columnwidth}{!}{\begin{tabular}{cSSSS}
\hline\hline
& \multicolumn{2}{c}{Tensorial} & \multicolumn{2}{c}{Watson} \\
\cline{2-5}
$K$ & \multicolumn{1}{c}{$\nu_{19}$} & \multicolumn{1}{c}{$\nu_{20}$} & \multicolumn{1}{c}{$\nu_{19}$} & \multicolumn{1}{c}{$\nu_{20}$} \\
\hline
\multirow{2}{*}{1}& -6.746 & 2.896 & -6.755 & 2.889 \\
    & 3925.582 & 1660.523 & 3919.740 & 1654.376 \\
\hline
\multirow{2}{*}{10}& -674.575 & 289.565 & -675.522 & 288.883 \\
    & 39255.816 & 16605.232 & 39197.407 & 16543.763 \\
\hline
\multirow{2}{*}{50} & -16864.375 & 7239.125 & -16888.050 & 7222.075 \\
& 196279.079 & 83026.158 & 195987.036 & 82718.816 \\
\hline
\multirow{2}{*}{80}& -43172.800 & 18532.160 & -43233.408 & 18488.512 \\
& 314046.526 & 132841.853 & 313579.258 & 132350.105 \\
\hline 
  \end{tabular}}
  \end{center}
\end{table}

The $l-$doubling constant $q_+$ well agrees between them for both $\nu_{19}$ and $\nu_{20}$ bands (1.4\% variation in average). Nevertheless, values given in the publication of Oka et \emph{al}.\cite{oka1964microwave} ($q_+(\nu_{19})=7.50$ and $q_+(\nu_{20})=12.60$) seem to be swapped with regards to this work. 

Finally, regarding the parameter $\eta_J$, that represent the $J(J+1)$ dependence of Coriolis coupling constant, we report a mismatch in the two $E$ bands that can be explained by the fact that higher order parameters are used in tensorial formalism; as a consequence, comparison has little meaning here.

\begin{landscape}
\renewcommand{\arraystretch}{1.25}
\begin{table}[ht]
 \caption{Spectroscopic constants of the ground-state and the three ground-vibrational state, $\nu_7$, $\nu_{19}$ and $\nu_{20}$ bands of Trioxane. These values were obtained from the tensorial formalism and converted to the Watsonian. As explained in Sect.~\ref{sec:watson_to_tensorial}, to apply the conversion formulas, ground state value is added to the excited state parameters ($t_k^i=\Delta t_k^i+t_0^k$).}
\label{tab:tensor-to-watson}
\begin{center}
\resizebox{\columnwidth}{!}{\begin{tabular}{
	l
	S
	S
	S
	S
	}
\hline\hline
\multicolumn{1}{c}{Parameters} & \multicolumn{1}{c}{\text{GS}} & \multicolumn{1}{c}{\text{$\nu_7$}}  & \multicolumn{1}{c}{\text{$\nu_{19}$}} & \multicolumn{1}{c}{\text{$\nu_{20}$}} \\
\hline
$T_v$ (cm$^{-1}$) &                 & 466.618966(37)& 524.466369(30)& 297.703777(50) \\
$B$ (MHz)     & 5273.257035(36) & 5260.2359(11) & 5276.2202(6) & 5270.2176(15) \\
$C$ (MHz)     & 2930.892885(36) & 2925.7498(15) & 2927.1103(10) & 2930.7491(24) \\
$D_J$ (kHz)   & 1.343880(11)    & 1.27589(23)   & 1.369534(43)  & 1.33740(17) \\
$D_{JK}$ (kHz)& -2.016314(48)   & -1.88681(16)  & -2.07951(49)  & -1.9950(19) \\
$D_K$ (kHz)   & 0.174643(41)    & 0.11306(51)   & 0.20769(41)   & 0.1616(14) \\
$\zeta$       &                 &               & -0.6705558(12)& -0.2832933(11) \\
$\eta_J$      &                 &                 &-3.9752(81) &-2.2402(45) \\
$q_+$         &                 &               & 13.0043(11) & 7.7873(22) \\
$H_J$ (mHz)   & 0.49044(76)     & 0.49044$^\dag$  & 0.49044$^\dag$  & 0.49044$^\dag$ \\
$H_{JK}$ (mHz)& -2.0962(21)     & -2.0962$^\dag$  & -2.0962$^\dag$  & -2.0962$^\dag$ \\
$H_{KJ}$ (mHz)& 2.7331(16)      & 2.7331$^\dag$   & 2.7331$^\dag$   & 2.7331$^\dag$ \\
\hline
\multicolumn{5}{l}{\footnotesize{$^\dag$ $\nu_7$, $\nu_{19}$ and $\nu_{20}$ sextic parameters are fixed to the ground-state values.}}
\end{tabular}}
\end{center}

\end{table}
\end{landscape}

\subsection{The first overtone \texorpdfstring{$2\nu_{20}$}{2v20}}\label{seq:overtone_issue}
\begin{figure*}[ht]
\centerline{\includegraphics[width=0.9\textwidth]{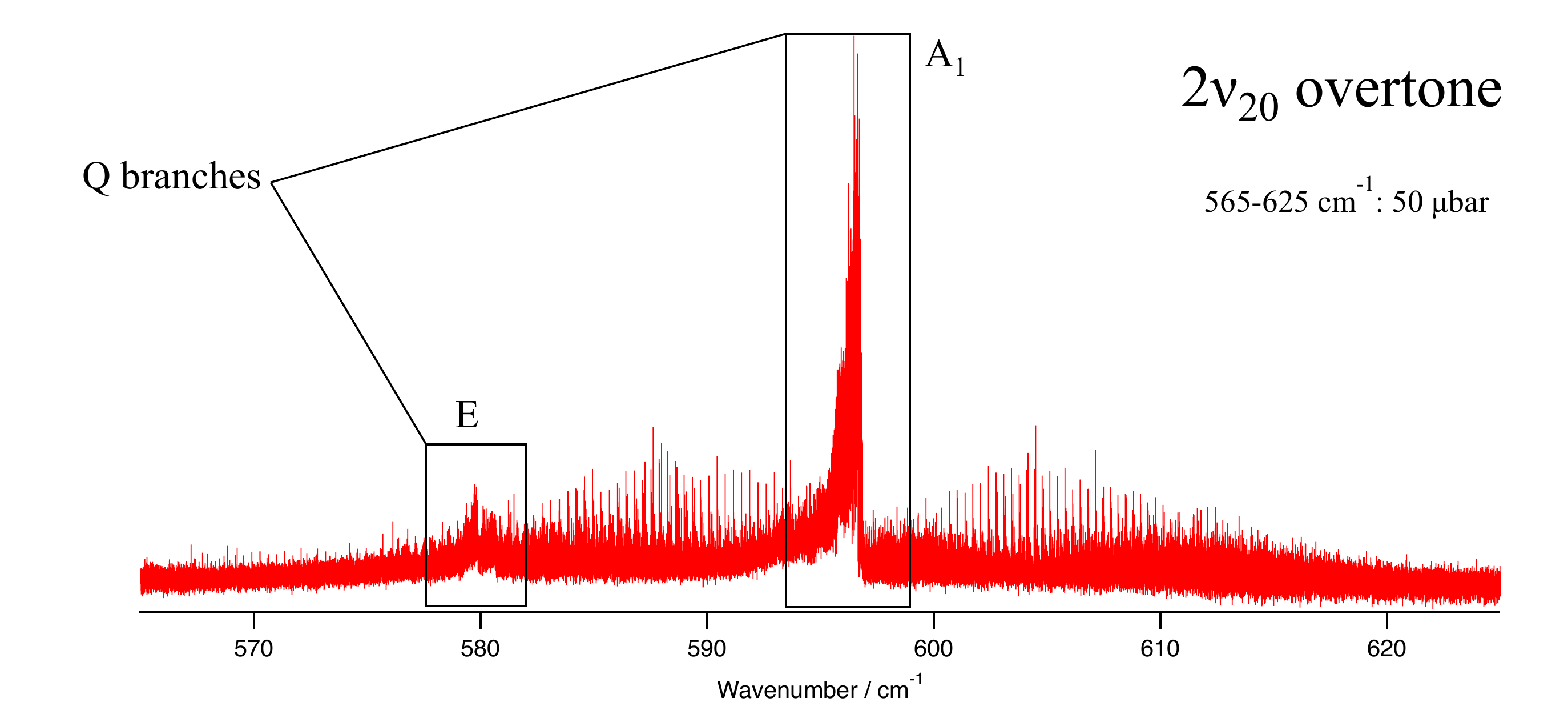}}
\caption{Part of the spectrum showing the first overtone $2\nu_{20}$ composed of a parallel band and a perpendicular band very close in energy. While the shape of the two $Q$ branches is visible, the perpendicular band has a poor signal-to-noise ratio and shows a very strong perturbation.}
\label{fig:overtone}
\end{figure*}

The $2\nu_{20}$ band, centered around \SI{595}{\wn}, is composed of a parallel sub-band ($A_1$) and a perpendicular sub-band ($E$) very close in energy as illustrated in Fig.~\ref{fig:overtone}. The $E$ sub-band is less intense but $Q$ branches are clearly visible for both sub-bands. Using parameters fitted for $\nu_{20}$ as a starting point of a new fit gives a prediction that seems close in a first look. Nevertheless, while line assignment in the $A_1$ sub-band seems to be not too difficult thanks to the fair enough signal intensity, it is not the same for the other.
 
\begin{figure*}[ht]
\centerline{\includegraphics[width=0.9\textwidth]{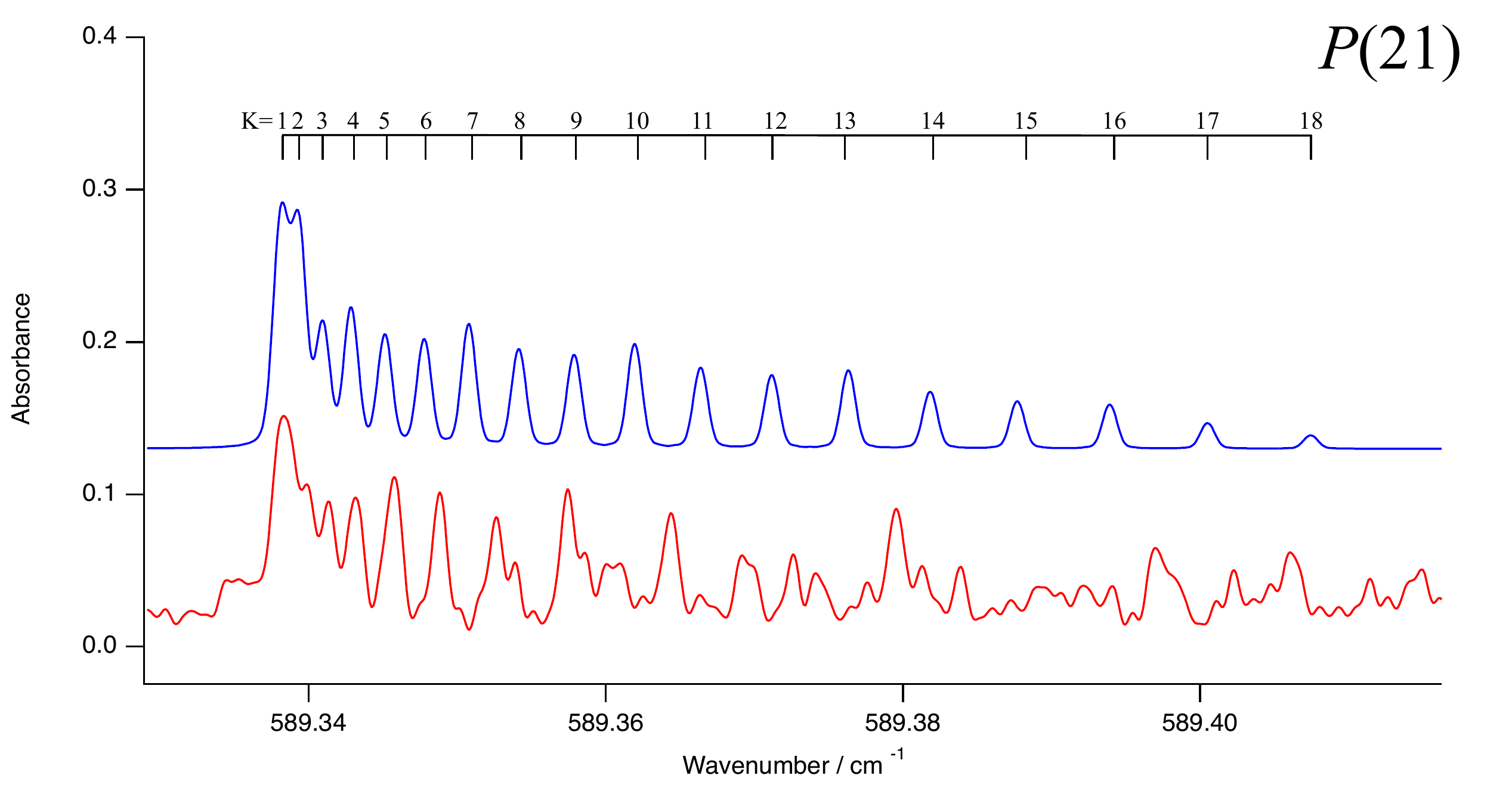}}
\caption{Zoom on the $K$ series of $P$(21) in the 2$\nu_{20}$ band. The higher the $K$, the more visible the perturbations become.}
\label{fig:2nu20_perturbation}
\end{figure*}

Attempts have been made to fit data iteratively as it was done for other bands. Nevertheless, even if the line pattern is clearly visible, perturbations quickly grow as $K$ value does, as illustrated in Fig.~\ref{fig:2nu20_perturbation}. Analysis of $A_1$- and $E$- level interaction is then needed but the signal-to-noise ratio is not good enough for the $E$ sub-band and assignment is not possible so far. As a consequence, $A_1$ sub-band cannot be analyzed in its entirety, especially for high $K$ values, since we cannot caracterise interaction with the $E$ sub-level. Recording another spectrum with a longer optical path could help to fix this issue. The analysis of the $2\nu_{20} \leftarrow \nu_{20}$ hot band, observed along with the $\nu_{20} \leftarrow 0$ fundamental transition could enable to assign unambiguously many $\nu_{20}=2$ rovibrational states and untangle this overtone analysis.

\section{Conclusion}

Ro-vibrational spectra were analysed in the laboratory for the three bands $\nu_{7}$, $\nu_{19}$ and $\nu_{20}$ bands of the trioxane molecule. We could determine with accuracy a set of spectroscopic constants derived from Watson's and tensorial formalism for $C_{3v}$ molecules developed in the Dijon group. This study allowed us to compare results from both models in a reliable way. From what we have observed, tensorial formalism is really well adapted to this type of molecule with such a symmetry. This model allows a systematic construction of all necessary operators up to any order of the development and this makes it easy to program and more flexible. Nevertheless, despite the fact we originally used this formalism to try to analyze the 2$\nu_{20}$ band, it did not work better. The spectrum of the first overtone of the $\nu_{20}$ mode was recorded but analysis could not get further because of the difficulty  in assigning lines in the perpendicular sub-band $E$. Assigning the $\nu_{20}=2$ rovibrational states from the $2\nu_{20}\leftarrow \nu_{20}$ hot band could be of great help for going further in the analysis of the first overtone.

\section*{Acknowledgments}
The authors are indebted to O. Pirali for the optical alignment and tuning of the long path absorption cell coupled to the high resolution interferometer. The authors are grateful to Soleil and the AILES staff for providing synchrotron beam under the proposal 20171389.

\bibliographystyle{elsarticle-num} 
\bibliography{biblio}

\end{document}